\documentclass[reprint,
 amsmath,amssymb,
prl,
]{revtex4-1}
\usepackage{xparse}

\usepackage{appendix}
\usepackage{graphicx}
\usepackage{dcolumn}
\usepackage{bm}

\renewcommand{\d}[2]{\ensuremath{\frac{\text{d} #1}{\text{d} #2}}}

\newcommand{\beginsupplement}{
\clearpage
\pagebreak
\setcounter{equation}{0}
\setcounter{figure}{0}
\setcounter{table}{0}
\setcounter{page}{1}
\makeatletter
\renewcommand{\theequation}{S\arabic{equation}}
\renewcommand{\thefigure}{S\arabic{figure}}
\onecolumngrid
\section{Supplementary Material for:\\ Kondo Impurity at the Edge  of a Superconducting Wire: Exact Results}
}
\newcommand{\ket}[1]{\ensuremath{\left| #1 \right>}}

\newcommand{\Tr}{\text{Tr}}
\usepackage{bbold}
\begin{document}

\title{Kondo impurity at the edge of a superconducting wire}

\author{Parameshwar R Pasnoori$^{\dagger}$, Colin Rylands$^{\ddagger}$ and  Natan Andrei} 
\email{prp56@physics.rutgers.edu}
\affiliation{$^{\dagger, *}$Department of Physics, Rutgers University, Piscataway, New Jersey 08854,\\ $^{\ddagger}$ Joint Quantum Institute and Condensed Matter Theory Center, Department of Physics,
University of Maryland, College Park, Maryland 20742-4111, U.S.A.}

\date{\today}

\begin{abstract}
Quantum impurity models are prevalent throughout many body physics, providing some prime examples of strongly correlated systems. Aside from being of great interest in themselves they can provide deep insight into the effects of strong correlations in general. The classic example is the Kondo model wherein a magnetic impurity is screened at low energies by a non interacting  metallic bath. Here we consider a magnetic impurity coupled to a quantum wire with pairing interaction which dynamically generates a mass gap. Using Bethe Ansatz we solve the system exactly finding that it exhibits both screened and unscreened phases for an antiferromagnetic impurity. We determine the ground state density of states and magnetization in both phases as well as the excitations. In contrast to the well studied case of magnetic impurities in superconductors we find that there are no intragap bound states in the spectrum. The phase transition is not associated to a level crossing but with quantum fluctuations.
\end{abstract}

\maketitle

\begin{figure}\centering
\includegraphics[trim={0cm 0cm 10cm 0cm},clip,width=.45\textwidth]{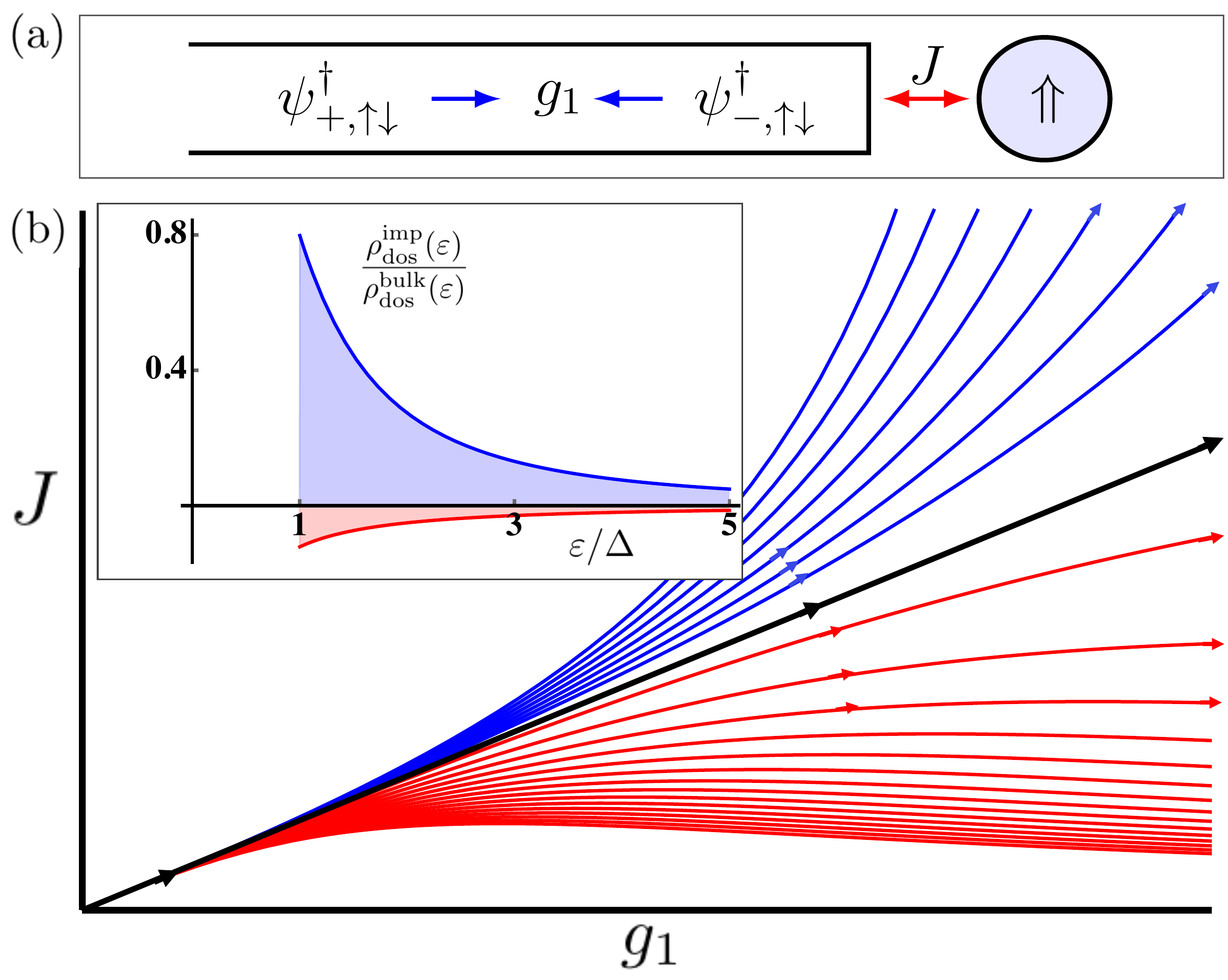}
\caption{Color Online. (a) A schematic of our system. A semi infinite, interacting quantum wire is coupled to a magnetic impurity at the boundary. The bulk interaction strength is $g_1$ and the coupling to the impurity is $J$.  (b) Main: The weak coupling RG flow diagram of our system, the flow is away from the non interacting point $J=g_1=0$. The diagonal line is $J=2g_1$ corresponding to $T_\text{K}/\Delta\approx0.32$. Lines above this (blue) correspond to the screened phase wherein the impurity flows to strong coupling. Below the diagonal the lines (red) correspond to the unscreened phase. We see that $J$ no longer flows toward strong coupling and no Kondo scale is generated. Inset: Impurity contribution to the density of states. The impurity enhances the density of states in the screened (blue) phase and suppresses it in the unscreened phase (red). }\label{Fig}
\end{figure}

\textit{Introduction.}\textemdash The basic quantum impurity model  describes a single magnetic impurity coupled to a metallic electron bath. The apparent simplicity of this model, the Kondo model, belies the strongly correlated physics it describes:  a dynamically generated energy scale $T_K$ and impurity screening at low energies and asymptotic freedom at high energies. The physics of the Kondo effect underpins our understanding of many disparate systems ranging from quantum dots to heavy fermion materials\cite{Hewson, ColemanPiers} and provides a proving ground for many powerful many-body techniques \cite{AndersonYuvalHamann, WilsonRMP, AndreiRMP, TWAKM, Affleck}. When the electrons in the bath interact among themselves the Kondo effect 
needs to be reexamined. A case of great interest is the interplay of the Kondo effect and superconductivity which has been intensely studied, both the impact of magnetic impurities on superconductivity \cite{Zitt,Aoi,Maple, LUENGO} and more recently, the impact of superconductivity  on the Kondo effect\cite{Franke, Zazaunov}. In this work we consider the latter issue, studying a system consisting of  a Kondo impurity placed at the edge of an attractively interacting quantum wire, see Figure \ref{Fig}. The attractive interactions among left and right moving electrons dynamically generate a superconducting mass gap $\Delta$, but as long range order is not allowed in one dimension \cite{MerminWagner}, the rigid phase correlations  and the charge sector decouple from the gapped spin sector \cite{Witten79, AndreiLowenstein79}.  Using Bethe Ansatz we solve exactly the model that describes the system and study its ground state properties through its density of states and magnetization.   We find that for an antiferromagnatic impurity and attractive bulk interactions the system exhibits two phases: a Kondo screened phase wherein the ground state is a spin singlet with odd fermion parity and an unscreened local moment  phase wherein the ground state is an even parity  spin doublet. The phase transition occurs at the ratio of Kondo temperature to mass gap of $T_K/\Delta\approx 0.32$.  This is reminiscent of a magnetic impurity in a 3-dimensional superconductor\cite{Sakurai,Shiba,Yu, Rusinov,  KondoSCRMP, KondoPRX,BotolinIucci}. However in contrast we find that there are no intragap bound states in the spectrum in either phase and so the phase transition is not due to a level crossing but to quantum fluctuations.

\textit{Hamiltonian.}\textemdash The Hamiltonian of our system is given by  $H=H_0+H_\text{int}+H_\text{imp}$. 
The first term, $H_0=\sum_{\sigma, a}\int_{-\frac{L}{2}}^{0}\mathrm{d}x\,\psi^{\dagger}_{\sigma,a}(x)\{-i\sigma\partial_x\}\psi_{\sigma,a}(x)$ is the  kinetic energy of the system, 
$\psi^\dag_{\sigma, a}(x)$ and $ \psi_{\sigma, a}(x) $ are the fermion creation and annihilation operators  with $\sigma=+,-$ indicating the chirality (right and left movers respectively), $a=\uparrow,\downarrow$ indicating the spin. We have set $\hbar,v_F=1$. The system is restricted to the half line with the boundary condition at $x=0$ imposed by taking $\psi_{+,a}(0)=\psi_{-,a}(0)$ and similarly at $x=-L/2$. 
The next term describes attractive spin  exchange interactions between fermions of opposite chirality,
$H_\text{int}=2g_1 \int_{-\frac{L}{2}}^{0}\mathrm{d}x \,\psi^{\dagger}_{+,a}(x)\psi_{+,b}(x)\psi^{\dagger}_{-,b}(x)\psi_{-,a}(x)$. 
To this we couple the impurity which lives on the right edge, $x=0$. The right moving electrons may scatter off the boundary, interact with the impurity via spin exchange and become left movers. This interaction is described by $H_\text{imp}=J  \vec{\sigma}_{ab}\cdot\vec{S}_{\alpha\beta}\psi^{\dagger}_{-,a}(0)\psi_{+,b}(0)$.
Here $\vec{\sigma}_{ab}$, $\vec{S}_{\alpha\beta}$ are Pauli matrices acting on the particle and impurity spin spaces respectively. 
  For $g_1=0$ the system reduces to the Kondo Hamiltonian whilst when $g_1\neq0$ but $J=0$ we recover the  Gross-Neveu Hamiltonian. In both limits the model can be solved exactly via Bethe Ansatz\cite{Andrei80, Wiegmann,  AndreiLowenstein79}. We show here that when both $J,g_1\neq 0$ the system can also be solved exactly, exhibiting different behavior that depends on the relative values of the interaction strengths. 
  
  \textit{Eigenstates.}\textemdash The Hamiltonian commutes with total particle number, $N=\int \psi_+^\dag(x)\psi_+(x)+\psi_-^\dag(x)\psi_-(x)$ and so we can diagonalize $H$ by constructing the exact eigenstates in each $N$ sector. The $N$-particle eigenstate takes the standard Bethe Ansatz form of a plane wave expansion in different regions of coordinate space. The state with energy $E=\sum_{j=1}^Nk_j$ is given by,
  \begin{eqnarray}\nonumber
\ket{\{k_j\}}=
\sum_{\substack{\{a_j\},\{\sigma_j\}\\Q}}\int \mathrm{d}\vec{x}\,F^{\{\sigma\}}_{\{a\}}(\vec{x})\prod_{j=1}^N \psi^{\dagger}_{\sigma_j,a_j}(x_j)\ket{0}
\otimes \ket{a_0}.
\end{eqnarray}
where $F^{\{\sigma\}}_{\{a\}}(\vec{x})=\theta(x_Q) A^{\{\sigma\}}_{\{a\}}[Q] e^{i\sum_{j}^N\sigma_j k_jx_j}$ is the $N$-particle wavefunction. 
Here the sum is over all different chirality and spin configurations specified by $\{\sigma_j\}=\{\sigma_1,\cdots ,\sigma_N\}$, $\{a_j\}=\{a_0,a_1,\cdots a_N\}$ where $\sigma_j=\pm$ and $a_j=\uparrow, \downarrow$ are the chirality and spin of the $j^\text{th}$ particle. $a_0$ denotes the spin of the impurity with $\ket{a_0}$ being the state of the impurity and $\ket{0}$ is the vacuum which contains no particles. We also sum over all different orderings of the particles, labelled by $Q$ which are elements of the symmetric group $Q\in \mathcal{S}_N$. $A^{\{\sigma\}}_{\{a\}}[Q]$ are the amplitudes for a particular spin and chirality configuration and  ordering of the particles while $\theta(x_Q)$ is a Heaviside function which is non zero only for the ordering of particles labelled by $Q$. These amplitudes are related to each other via application of the various $S$-matrices in the model which are determined by the $N$-particle Schr\"{o}dinger equation and the consistency of the solution. The $S$-matrices are given by \cite{Supplement}
\begin{eqnarray}\label{S12}S^{ij}= \frac{2ib \; I^{ij} +P^{ij}}{2ib+1},~
S^{j0}=\frac{\; I^{j0} -i cP^{j0}}{1-ic}\end{eqnarray}
where $I^{ij}$ is the identity operator
 and $P^{ij}=(I^{ij}+ \vec{\sigma}_i\cdot\vec{\sigma}_j)/2$ is the permutation operator acting on the spin spaces of particles $i$ and $j$ with $0$ indicating the impurity. Furthermore we have introduced the parameters $b= \frac{1-g_1^2/4}{2g_1}$ and  $c=\frac{2J}{1-3J^2/4}$ which encode separately the bulk and impurity coupling constants. $S^{j0}$ is the impurity $S$-matrix describing the interaction of the electrons with the impurity. It relates amplitudes which  differ by   changing the chirality of the right most particle $+\to -$. Similarly, amplitudes which are related by swapping the order of particles with different chiralities are related by the particle-particle $S$-matrix, $S^{ij}$. An additional  $S$-matrix, denoted by $W^{ij}$, is also required. It  relates amplitudes that differ by exchanging particles of the same chirality. This is given by $W^{ij}=P^{ij}$. The consistency of the solution is then guaranteed as the $S$-matrices satisfy the Yang-Baxter and Reflection equations
 \cite{Sklyannin, Cherednik, ZinnJustin}.

Imposing the boundary condition at $x=-L/2$ quantizes the single particle momenta $k_j$ and allows us to determine the spectrum of $H$. We find \cite{Supplement},
\begin{eqnarray}\label{energy}
e^{-ik_jL}\!=\!\prod_{\alpha=1}^Mf(2bc, 2\lambda_\alpha),~
f(x,z)=\!\prod_{\sigma=\pm}\frac{x+\sigma z+ic}{x+\sigma z-ic}
\end{eqnarray}
where the newly introduced $\lambda_\alpha$, $\alpha=1,\dots,M$ are the Bethe roots which satisfy the Bethe Ansatz equations
\begin{eqnarray}\label{BAE}
\left[f(2\lambda_\alpha,2bc)\right]^{N}f(2\lambda_j,2dc)=\prod_{\alpha\neq \beta }^Mf(\lambda_\alpha,\lambda_\beta)
.
\end{eqnarray}
 with $d=\sqrt{b^2-2b/c-1}$. The Bethe roots govern the spin degrees of freedom of the system while $M$ is related to the total $z$ component of spin. The solutions of \eqref{BAE} allow for $\lambda_\alpha$ to be real or take complex values in the form of strings\cite{Takahashi} and in order to have a non vanishing wavefunction they must all be distinct, $\lambda_\alpha\neq \lambda_\beta$. In addition, the value $\lambda_\alpha=0$ will also result in a vanishing wavefunction \cite{ODBA} and so any solution containing this root should be removed.

\textit{The Ground state.}\textemdash 
The model exhibits several different phases depending on the values of $b$, $c$ and  $d$. In the absence of the impurity, $b>0$ corresponds to the gapped  phase  of the model  while on the other hand when there are no bulk interactions but the impurity is present $c>0$ corresponds to the Kondo regime.  We consider here only $b,c>0$ which allows us to study the interplay of these two effects. Within this regime $d$ can be either real or imaginary with the structure of the Bethe equations dependent on this. We will see below that real and imaginary $d$ correspond to two different phases of the model and the ground state needs to constructed separately in each case. 

 We begin by constructing the ground state for $d$ being real which for reasons which will become evident we refer to as the screened phase. In the ground state all Bethe roots are real \cite{AndreiLowenstein79}. Taking the logarithm of \eqref{BAE} we have that the Bethe equations become

\begin{eqnarray}\nonumber
\sum_{\sigma=\pm} N\Theta(\lambda_\alpha+\sigma bc,c/2)+\Theta(\lambda_\alpha+\sigma dc ,c/2)\\+\Theta(\lambda_\alpha,c/2) \label{Logbae}
=\sum_{\beta=1}^{M} \sum_{\sigma=\pm}\Theta\left(\lambda_\alpha+\sigma \lambda_\beta,c\right)+\pi I_j\end{eqnarray}

Where $\Theta(x,n)=\text{arctan}[x/n]$. And likewise taking the logarithm of \eqref{energy} we get
\begin{eqnarray}\label{logEnergy}
k_j=\frac{2\pi n_j}{L}+\frac{2}{L}\sum_{\beta=1}^M\sum_{\sigma=\pm}\Theta( bc +\sigma\lambda_\beta,c/2)
\end{eqnarray}

Here  $n_j$ and $I_j$ are integers which arise from the logarithmic branch, they serve as the quantum numbers of the system. The  quantum numbers $n_j$ are associated with the charge degrees of freedom 
are arbitrary implying their decoupling from the spin sector. They require the imposition of a cutoff such that $2\pi|n_j|/L < \pi D$ where $D=2N/L$ is the density. Similarly the quantum numbers $I_\alpha$ correspond to the spin degrees of freedom, and any allowed choice of these quantum numbers correspond to an eigenstate. The ground state is given by the choice  $n^0_j$ , $I^0_\alpha$ where $n^0_j$ are consecutively filled from the cutoff up and the integers $I^0_\alpha$ also take consecutive values which corresponds to taking only real values of $\lambda_\alpha$. 

The thermodynamic limit entails taking $N,L \rightarrow \infty$, with $D$ held fixed. In this limit the Bethe roots fill the real line and the ground state can be described by a distribution, $\rho_\text{s}(\lambda)$ from which the properties of the ground state can be obtained. This ground state distribution is determined by the following integral equation which is derived from \eqref{Logbae}\cite{Supplement},

\begin{eqnarray}\label{gsdensity}
g_s(\lambda)&=&\rho_\text{s}(\lambda)+\int\mathrm{d}\mu\,\varphi(\lambda-\mu,c)\rho_\text{s}(\mu)
\end{eqnarray}

where $g_s(\lambda)=\sum_{\sigma=\pm}N\varphi(2\lambda+2\sigma bc ,c) +\varphi(2\lambda+2\sigma dc ,c)+\varphi(2\lambda,c)-(1/2)\delta(\lambda)$ and  $\varphi(x,n)= (n/\pi)(n^2+x^2)^{-1}.$


Solving \eqref{gsdensity} by Fourier transformation  we get
\begin{equation}\label{scr}\tilde\rho_\text{s}(\omega)= \frac{2N \cos[ bc\,\omega] + 2\cos[d c\,\omega]+ 1-e^{\frac{c|\omega|}{2}}}{4\sqrt{2\pi}\cosh[\frac{c\omega }{2}]}. \end{equation}
 Each of the terms here may be identified with a certain component of the system. The term which is proportional to $N$ is the contribution of the left and right moving fermions of the quantum wire, the next term which depends upon $d$ is the contribution due to the impurity while the remaining terms can be associated with the boundaries at $x=0,-L/2$. From the density $\rho_\text{s}(\lambda)$ we obtain 
the total $z$-component of spin in the ground state,
$\left<S_z\right>=(N+1)/2-\int_{-\infty}^{\infty}\mathrm{d}\lambda\, \rho(\lambda).$ 
Using $\sqrt{2\pi} \tilde\rho(0)=\int\mathrm{d}\lambda\, \rho(\lambda)$ along with \eqref{scr} we find that $\left<S_z\right>=0$ meaning that the ground state is a spin singlet. In addition the parity is given by $(-1)^N=-1$. Therefore the impurity spin has been completely screened by the electrons in the wire, indicative of the Kondo effect. To confirm this we need to compute the density of states and magnetization which we shall do below. 

Considering now $d= ia$ to be imaginary we shall find the ground state in the unscreened phase. Repeating the same analysis\cite{Supplement}  the ground state distribution of the Bethe roots turns out to be,
\begin{equation}\label{unscr}\tilde\rho_\text{us}(\omega)=\frac{2N \cos[ bc\, \omega]+e^{-ac|\omega|}-e^{-( a-1)c|\omega|}+1-e^{\frac{c|\omega|}{2}}}{4\sqrt{2\pi}\cosh[\frac{c\omega}{2}]}.  \end{equation}
We  note here that the  bulk and boundary terms are the same as in the screened phase however the impurity term is different. This difference carries over to the total spin which can be calculated as before,  $\left<S_z\right>=1/2$ meaning that there is an unscreened spin half in the ground state of the system. Furthermore the parity is $(-1)^N=1$.  The ground state in the unscreened phase is therefore an even parity doublet. 

\textit{Excitations.}\textemdash In order to confirm that we have indeed found the true ground state of the model in the screened and unscreened phases we investigate other solutions to the Bethe equations to check that they correspond to excitations which increase the energy.
 Excitations arise from modifying the quantum numbers, $n_j$ or $I_\alpha$. Note that we can choose $n_j$ and $I_\alpha$ independently, meaning that the spin and charge degrees of freedom are decoupled\cite{Haldane, Witten79, AndreiLowenstein79}. In the charge sector the excitations are constructed by removing a number, $n^h<0$ from the sequence $n^0_j$ and adding an extra $n^p>0$. The energy of this excitation is $\delta E=2\pi(n^p-n^h)/L>0$. Gapless excitations such as this are known as holons.
The structure of excitations in the spin sector is more complicated as they arise from alternative solutions to the Bethe Ansatz equations \eqref{BAE}.  In the supplementary material we provide a detailed analysis of possible low energy excitations of the models following the method of Destri and Lowenstein \cite{DestriLowenstein, Supplement}. We find that the lowest energy excitations are of two spinons which form either a triplet or singlet. The triplet excitation is constructed by removing two, arbitrary Bethe roots, $\lambda^h_1,\lambda^h_2$ from the ground state distribution \cite{AndreiLowenstein79}. Each hole corresponds to a single spinon with spin $+1/2$. In order to form a singlet one must add to this a complex conjugate pair of Bethe roots at $\lambda_s\pm ic/2$ where $\lambda_s=\sqrt{([\lambda^h_1]^2+[\lambda^h_2]^2)/2+c^2/4}$ \cite{GrisaruMezincescNepomechie}. The spin-1/2 spinons can thus be in total spin state $S=1,0$ respectively while in both cases the energy is \cite{Supplement},
\begin{eqnarray}\label{tripletenergy}\delta E=  \sum_{l=1}^2D\arctan\left[\frac{\cosh(\pi \lambda^h_l/c)}{\sinh\left(b\pi\right)}\right].\end{eqnarray}
From this we find that the system has dynamically generated a superconducting mass gap in the spin sector  $\Delta=D\arctan{[\sinh(\pi b)]^{-1}}$. 

Up to now we have kept a finite cutoff, however to obtain universal answers we need to take the scaling limit, $D\to\infty$ while holding the physical mass $\Delta$ fixed. In this limit we have that $\Delta=2De^{-\pi b}$ and the excitation energy of a single spinon becomes  $\varepsilon(\lambda)=\Delta\cosh{(\lambda)}$ where we have absorbed $\pi/c$ into $\lambda$.
These spinons and holons are  bulk excitations and are the same in both the screened and unscreened phases. Furthermore it   can be shown that there are no lower lying excitations, in particular there are no intragap bound states \cite{Supplement}. This stands in contrast to the well studied case of  a magnetic impurity coupled to the BCS model where such bound states do exist\cite{Sakurai,Yu, Shiba, Rusinov}\footnote{we note however that in our case the superconductivity as well as the impurity are fully quantum fluctuating as opposed to the case studied in \cite{Sakurai,Yu, Shiba, Rusinov}}.

 \textit{Density of States.}\textemdash We have shown that the model exhibits two different phases which corresponds to a screened and unscreened spin $1/2$ in the ground state. To study this further we calculate the ground state density of states in both phases. This is given by $\rho_{\text{dos}}(\varepsilon)=|\rho_\text{gs}(\lambda)/\varepsilon'(\lambda)|$ \cite{4Lectures}. Where by  $\rho_\text{gs}$ we mean either the distribution of Bethe roots given in \eqref{scr} or \eqref{unscr}. This naturally separates into contributions from the bulk, boundary and impurity with the former two being the same in both phases, $\rho_\text{dos}=L\rho^\text{bulk}_\text{dos}+\rho^\text{bdry}_\text{dos}+\rho^\text{imp}_\text{dos}$. The bulk  contribution, per unit length is $\rho^\text{bulk}_\text{dos}(\varepsilon)=\frac{\varepsilon /\pi}{\sqrt{\varepsilon^2+\Delta^2}}$ with $\varepsilon\geq \Delta$. The contribution from the boundary is given by
 \begin{eqnarray}\nonumber
 \frac{\rho^{\text{bdry}}_\text{dos}(\varepsilon)}{\rho^\text{bulk}_\text{dos}(\varepsilon)}=\frac{\Delta}{4\varepsilon^2}-\frac{1}{4\pi\epsilon }\mathfrak{R}\!\!\left[\Psi\!\left(\frac{i\lambda+2}{2}\right)\!-\!\Psi\!\left(\frac{i\lambda+1}{2}\right)\right]
 \end{eqnarray}
where $\Psi$ is the digamma function  and $\lambda(\varepsilon)=\text{arccosh}(\varepsilon/\Delta)$.  In the screened phase we find  an enhancement of the density of states due to the impurity in agreement with well known Kondo peak \cite{Hewson}. The impurity contribution is given by
 \begin{eqnarray}
  \frac{\rho^{\text{imp}}_\text{dos}(\varepsilon)}{\rho^\text{bulk}_\text{dos}(\varepsilon)}=\frac{\Delta\cosh\left[\log{\left(\frac{4T_0}{\Delta}\right)}\right]}{2\varepsilon^2-\Delta^2+\Delta^2\cosh\left[\log{\left(\frac{4T_0}{\Delta}\right)^2}\right]}
 \end{eqnarray}
where we have introduced $2T_0=De^{-\pi/c}$ which is the strong coupling Kondo scale \cite{AndreiLowensein81}. The  form of this scale is chosen so that in $\Delta\to  0$ limit we recover the well known Lorentzian density of states of the Kondo model\cite{AndreiRMP}. At finite $\Delta$ this density of states also exhibits a Kondo peak at $\epsilon=\Delta$, see Figure \ref{Fig}. In the scaling limit we can identify $e^{\pi d}=4T_0/\Delta$ meaning the transition from the screened to the unscreened phase occurs at $T_0/\Delta=1/4$. The strong coupling Kondo scale can be related to the weak coupling scale known as the Kondo temperature $T_\text{K}=WT_0$ where $W\approx 1.29$ is Wilson's number \cite{WilsonRMP, AndreiRMP}. In terms of this we have that the system is in a  screened phase for $T_\text{K}/\Delta > 0.32$ with a phase transition occurring at $T_\text{K}/\Delta \approx 0.32$.  

In the unscreened phase we find that the find that the contribution of the impurity to the density of states is  instead given by \begin{eqnarray}\nonumber
 \frac{\rho^{\text{imp}}_\text{dos}(\varepsilon)}{\rho^\text{bulk}_\text{dos}(\varepsilon)}=\frac{1}{4\pi\epsilon }\mathfrak{R}\!\!\left[\Psi\!\left(\frac{i\lambda+a}{2}+\frac{1}{4}\right)\!+\!\Psi\!\left(\frac{i\lambda+a}{2}+\frac{3}{4}\right)\right.\\\left.-\Psi\!\left(\frac{i\lambda+a}{2}+\frac{1}{2}\right)\!-\!\Psi\!\left(\frac{i\lambda+a}{2}\right)
 \right]~~~
\end{eqnarray}
where in the scaling limit $a>1$ which is an RG invariant. This contribution is negative and so in the unscreened phase the impurity contributes to a suppression of the density of states, in line with the fact that there is an unscreened spin $1/2$ in the ground state. 

\textit{Magnetization.}\textemdash We may couple globally an external magnetic field to the system via the addition of $-\mu h S_z$ term to the Hamiltonian. This term is minimized by breaking pairs creating triplet excitations in the system which come at energy cost of at least $2\Delta$. For $h<2\Delta/\mu$ the ground state is unchanged from the expressions given previously, whereas for a large field, $h>2\Delta/\mu$,
the system will become magnetized and the ground state is changed. Re-solving for the ground state distribution in the screened phase we find that the total magnetization, for small $h-2\Delta/\mu$, is
\begin{eqnarray}\label{magnetscr}
 \left<S_z\right>= \frac{1}{2\sqrt{2}\pi} \left(\Delta L  +1 + \frac{4}{\left(\frac{\Delta}{4T_0}+\frac{4T_0}{\Delta}\right)}\right)\left(\frac{\mu h-2\Delta}{2\Delta}\right)^{1/2}.
\end{eqnarray}
The first term here is the magnetization of the bulk which agrees with the calculation in the Gross-Neveu model\cite{BahderRezayiSak}, the next term is due to the presence of the boundary while the remainder is the due to the impurity. We see here that the magnetization vanishes at $h=2\Delta/\mu$ and shows a square root dependence on the magnetic field otherwise. Note that this is in contrast to standard Kondo model in which the magnetisation is linear in the field\cite{Hewson}.

In the unscreened phase we find instead that 
\begin{eqnarray}\label{magnetunscr}
\left<S_{z}\right>=\frac{1}{2}+\frac{1}{2\sqrt{2}\pi} \left(\Delta L +1 + \frac{\mathcal{I}(a)}{\pi}\right)\left(\frac{\mu h-2\Delta}{2\Delta}\right)^{1/2}
\end{eqnarray}
where
$\mathcal{I}(a)= \Psi\left(\frac{a}{2}-\frac{1}{4}\right)+\Psi\left(\frac{a}{2}+\frac{3}{4}\right)-2\Psi\left(\frac{a}{2}+\frac{1}{4}\right)$. We see here the unscreened spin contribution as well as a modified contribution from the impurity, the $\mathcal{I}(a)$ term which provides a negative contribution to the magnetization. The impurity thus provides a negative susceptibility. This can be understood from the fact that the unscreened spin acts as a local magnetic field with negative magnetic moment, due to the antiferromegnetic interactions,  thereby reducing the local magnetization. 

\textit{Renormalization Group.}\textemdash We now express our findings in the language of the renormalization group. Using our solution we can drive the weak coupling RG equations \cite{Supplement},
\begin{eqnarray}\label{RG}
\d{J}{l}=-\frac{2}{\pi}J(J-g_1),~\d{g_1}{l}=-\frac{2}{\pi} g_1^2.
\end{eqnarray}
where $l=\log{\Lambda}$ is some energy scale with $\Lambda>\Delta$. These reduce to a single equation when $J=2g_1$ with the flow changing either side of this, the flow diagram is plotted in Figure \ref{Fig}. We see that within the screened phase $J$ flows to strong coupling whereas this is not the case in the unscreened phase. 

We can compare this to the well studied case of a magnetic impurity coupled to the BCS model. Therein also a phase transition occurs between an odd parity, singlet ground state and an even parity doublet. However, in that case the phase transition is first order and is the result of the presence of intragap bound states. In our model such states do not occur and instead the phase transition is driven by quantum fluctuations.

\textit{Acknowledgements.}\textemdash
This research was supported by the Rutgers Samuel Marateck Fellowship (PRP) DOE-BES (DESC0001911) (CR) an by NSF Grant DMR 1410583 (NA). We thank P. Coleman, J. Pixley, Y. Komijani, A. Culver and A. Iucci for stimulating discussions.



\bibliography{mybib}

\beginsupplement
\subsection{Construction of $N$-Particle Eigenstates}
In this section we show how to construct the eigenstates of $H$.
Since $N$ is a good quantum number we may construct the eigenstates by examining the different $N$ particle sectors separately. We start with $N=1$ wherein we can write the wavefunction as an expansion in plane waves,
\begin{eqnarray}\nonumber
\ket{k}=\sum_{a_j=\uparrow\downarrow,\sigma=\pm}\int_{-\frac{L}{2}}^{0}\mathrm{d}x\, e^{i\sigma kx} A^\sigma_{a_1a_0}    \psi^{\dagger}_{\sigma,a_1} (x)\ket{0} \otimes \ket{a_0}.
\end{eqnarray}  $\ket{0}$ is the drained Fermi sea and $A^\sigma_{a_1a_0}$  are the amplitudes for an electron with chirality $\sigma$ and spin $a_1$ and the impurity having spin $a_0$. These amplitudes are fixed by single particle Schrodinger equation.   Applying the Hamiltonian to $\ket{k}$ we find that it is an eigenstate of energy $k$ provided  $A^-_{a_1a_0}=S^{10}_{a_1b_1, a_0b_0} \; A^+_{b_1b_0}$ where $S^{10}_{a_1b_1, a_0b_0}$ is the impurity-particle S-matrix given by,
\begin{eqnarray}\label{Simp}
S^{10}_{a_1b_1,a_0b_0}=e^{i\gamma} \; \left(\frac{\; I^{10}_{a_1b_1,a_0b_0}-icP^{10}_{a_1b_1,a_0b_0}}{1-ic}\right),\\\label{c} c=\frac{2J}{1-3J^2/4},~e^{i\gamma}=\frac{2iJ-1+3J^2/4}{iJ-1-3J^2/4}.\end{eqnarray}
 Here $P^{10}$ is the usual permutation operator $P^{10}_{a_1b_1,a_0b_0}=(I^{10}_{a_1b_1,a_0b_0}+ \vec{\sigma}^1_{a_1b_1}\cdot\vec{\sigma}^0_{a_0b_0})/2$ which exchanges the spins of particle and impurity. The superscripts refer to the particle space on which this operator acts i.e. $1$ refers to the particle and $0$ to the impurity. This completes the construction of eigenstates for $N=1$.

We next consider the two particle sector, $N=2$, were $H_\text{int}$ plays a role.
Since the two particle interaction is point-like
we may divide configuration space into regions such that
the interactions only occur at the boundary between two
regions. Therefore away from these boundaries we write
the wavefunction as a sum over plane waves so that the most general two particle state can be written
\begin{eqnarray}\label{2particle}
\ket{k_1,k_2}&=& \sum_{\sigma,a} \int_{-\frac{L}{2}}^{0}\mathrm{d}^2x\,F_{a_1a_2a_0}^{\sigma_1\sigma_2}(x_1,x_2)e^{\sum_{j=1}^2i\sigma_jk_jx_j}\psi^{\dagger}_{\sigma_1a_1}(x_1)\psi^{\dagger}_{\sigma_2a_2}(x_2) \ket{0}\otimes\ket{a_0},
\end{eqnarray} 
where we sum over all possible spin and chirality configurations  and the two particle wavefunction, $F_{a_1a_2a_0}^{\sigma_1\sigma_2}(x_1,x_2)$ is split up according to the ordering of the particles,
\begin{eqnarray}\nonumber
 F_{a_1a_2a_0}^{\sigma_1\sigma_2}=A_{a_1a_2a_0}^{\sigma_1\sigma_2}[12]\theta(x_2-x_1)+A_{a_1a_2a_0}^{\sigma_1\sigma_2}[21]\theta(x_1-x_2).
\end{eqnarray}
The amplitudes $A_{a_1a_2a_0}^{\sigma_1\sigma_2}[Q]$ refer to a certain spin and chirality configuration, specified by $\sigma_j$, $a_j$ as well as an ordering of the particles in configuration space denoted by $Q$. For $Q=12$ particle $1$ is to the left of particle $2$ while for $Q=21$ the  order of the particles are exchanged.
Applying the Hamiltonian to \eqref{2particle} we find that it is an eigenstate with energy $E=k_1+k_2$ provided that these amplitudes are related to each other via application of  $S$-matrices. As in the single particle case, amplitudes which differ changing the chirality of a particle are related by the impurity-particle $S$-matrix,
\begin{eqnarray}
A^{\sigma_1-}[12]=S^{20}A^{\sigma_1+}[12],\\
A^{-\sigma_2}[21]=S^{10}A^{+\sigma_2}[21]
\end{eqnarray}
 where $S^{j0}$ is given by \eqref{Simp} acting on the $j^\text{th}$ particle space and for ease of notation  we have suppressed spin indices.

In addition there are two types of two particle $S$-matrices denoted by $S^{12}$ and $W^{12}$ which arise due to the bulk interactions and relate amplitudes which have different orderings. The first relates amplitudes which differ by exchanging the order of particles with opposite chirality 
\begin{eqnarray}
A^{+-}[21]=S^{12}A^{+-}[12]\\
A^{-+}[12]=S^{12}A^{-+}[21]
\end{eqnarray}
where  $S^{12}$ acts on the spin spaces of particles 1 and 2. Restoring the indices it is given by,
\begin{eqnarray}\label{S12}S^{12}_{a_1a_2,b_1b_2}= e^{i\phi}\; \left(\frac{2ib \; I^{12}_{a_1a_2,b_1b_2} +P^{12}_{a_1a_2,b_1b_2}}{2ib+1}\right),~~~\\\label{b}
b= \frac{4+(g_2^2-g_1^2)}{8g_1},~e^{i\phi}=\frac{1+(g_2^2-g_1^2)/4-ig_1}{1-(g_2^2-g_1^2)/4+ig_2}. \end{eqnarray}
Whilst the second relates amplitudes where  particles of the same chirality are exchanged,
\begin{eqnarray}
A^{--}[21]=W^{12}A^{--}[12],\\\label{W12}
A^{++}[12]=W^{12}A^{++}[21].
\end{eqnarray}
Once again the superscripts indicate the spin spaces on which $W^{12}$ acts. Unlike \eqref{S12}, $W^{12}$ is not fixed by the Hamiltonian but rather by the consistency of the construction. This is expressed through the reflection equation \cite{Sklyannin, Cherednik}
\begin{eqnarray}\label{RE}
S^{10}S^{12}S^{20}W^{12}=W^{12}S^{20}S^{12}S^{10}
\end{eqnarray}
which needs to be satisfied for the eigenstate to be consistent. We take $W^{12}=P^{12}$ which can be explicitly checked to satisfy \eqref{RE}. The relations \eqref{2particle}-\eqref{W12} provide a complete set of solutions of the two particle problem. 

We can now generalise this to the $N$-particle sector and find that the eigenstates of energy $E=\sum_{j=1}^Nk_j$ are of the form
\begin{eqnarray}\label{NparticleS}
\ket{\{k_j\}}=
\sum_{Q,\vec{a},\vec{\sigma}}\int \theta(x_Q) A^{\{\sigma\}}_{\{a\}}[Q] \prod_j^N e^{i\sigma_j k_jx_j}\psi^{\dagger}_{a_j\sigma_j}(x_j)\ket{0}
\otimes \ket{a_0}.
\end{eqnarray}
Here we sum over all  spin and chirality configurations specified by $\{a\}=\{a_1\dots a_Na_0\}$, $\{\sigma\}=\{\sigma_1\dots \sigma_N\}$ as well as different different orderings of the $N$ particles. These different orderings correspond to elements of the symmetric group $Q\in \mathcal{S}_N$. In addition $\theta(x_Q)$ is the Heaviside function which is nonzero only for that particular ordering. 
As in the $N=1,2$ sectors the amplitudes $A^{\vec{\sigma}}_{\vec{a}}[Q]$ are related to each other by the various $S$-matrices in the same manner as before i.e. amplitudes which differ by changing the chirality of the rightmost particle are related by the impurity $S$-matrix, $S^{j0}$, amplitudes which differ by exchanging the order of opposite or same chirality particles are related by $S^{ij}$ and $W^{ij}$ respectively. The consistency of this construction is then guaranteed by virtue of these $S$-matrices satisfying the following  reflection and  Yang-Baxter equations\cite{Sklyannin, Cherednik, ZinnJustin}
\begin{eqnarray}\label{YB1}
W^{jk} \;W^{ik}\; W^{ij} &=& W^{ij} \;W^{ik} \;W^{jk}\\\label{YB2}
S^{jk}\;S^{ik}\;W^{ij} &=& W^{ij}\;S^{ik}\;S^{jk}\\\label{YB3}
S^{j0}\;S^{ij}\;S^{i0}\;W^{ij} &=& W^{ij}\;S^{i0}\;S^{ij}\;S^{j0}
\end{eqnarray}
Where $W^{ij}=P^{ij}$ and as before the superscripts denote which particles the operators act upon.

\subsection{Bethe Equations}
In this section we derive the Bethe equations \eqref{BAE}. 
Enforcing the boundary condition at $x=-L/2$ on the eigenstate \eqref{NparticleS} we obtain the following eigenvalue problem which constrains the $k_j$,
\begin{eqnarray}
e^{-ik_jL}A^{\{\sigma\}}_{\{a\}}[\mathbb{1}]=\left(Z_j\right)^{\{\sigma\},\{\sigma\}'}_{\{a\},\{a\}'} A[\mathbb{1}]^{\vec{\sigma}'}_{\vec{a}'}.
\end{eqnarray}
Here $\mathbb{1}$ denotes the identity element of $\mathcal{S}_N$, i.e. $\mathbb{1}=12\dots N$ and the operator $Z_j$ is the transfer matrix for the $j^\text{th}$ particle given by
\begin{eqnarray}
Z_j=W^{jj-1}\dots W^{j1} S^{j1}...S^{jj-1}S^{jj+1}...S^{jN} S^{j0}W^{jN}...W^{jj+1}\end{eqnarray}
where the spin indices have been suppressed. This operator takes the $j^\text{th}$ particle from one side of the system to the other and back again, picking up $S$-matrix factors along the way as it moves past the other $N-1$ particles, first as a right mover and then as a left mover.   Using the relations \eqref{YB1} \eqref{YB2} and \eqref{YB3}, one can prove that all the transfer matrices commute, $[Z_j,Z_k]=0$ and therefore are simultaneously diagonalizable. In order to determine the spectrum of $H$ we must therefore diagonalize $Z_j,~\forall j$. Here we choose to diagonalize $Z_1$. To do this we use the method of boundary algebraic Bethe Ansatz \cite{Sklyannin, Cherednik, ODBA}.  In order to use this method we need to embed the bare S-matrices in a continuum that is, we need to find the matrices $R(\lambda)$, $K(\lambda)$ such that for certain values of the spectral parameter $\lambda$, we obtain the bare S-matrices of our model.  $R(\lambda)$ turns out to be the $R$-matrix of the \small{$XXX$} spin chain which is given by \small{\begin{equation}R^{ij}_{ab}(\lambda)=\frac{1}{i\lambda+1}\left(i\lambda I^{ij}_{ab} + P^{ij}_{ab}\right).\end{equation}} We can see that \small{$R^{ij}(0)=W^{ij}, \hspace{2mm} R^{ij}(2b)= S_{ij}$}. We have ignored the unimportant constant $e^{i\phi}$. The $K$-matrix is 
\small{$K^{j0}(\lambda)=R^{j0}(b+d)R^{j0}(b-d)=S^{j0}, \hspace{2mm} d= \sqrt{b^2-2b/c-1}$}. These $R$ and $K$ matrices form the Reflection algebra \cite{ODBA}. The transfer matrix is related to the Monodromy matrix $\Xi_{\tau}(\lambda)$ as $Z_1=t(b)=\Tr_{\tau} \Xi_{\tau}(b)$, where
\begin{equation}\Xi_{\tau}(\lambda)= R_{1\tau}(\lambda+b)...R_{N\tau}(\lambda+b)R_{0\tau}(\lambda+d)R_{0\tau}(\lambda-d)R_{N\tau}(\lambda-b)...R_{1\tau}(\lambda-b)\end{equation}
Here $\tau$ represents auxiliary space and $\Tr_\tau$ represents the trace in the auxiliary space.
Using the Reflections equations of the Reflection algebra and the properties of the $R$ matrices, one can prove that $[t(\lambda),t(\mu)]=0$. \cite{ODBA} . By expanding $t(\mu)$ in powers of $\mu$, one can obtain infinite set of conserved charges which guarantees integrability. By following the Boundary Algebraic Bethe Ansatz approach \cite{ODBA}, we obtain the Bethe equations 
\begin{equation}\label{Bae}
\left(\frac{\lambda-\tilde b+ic/2}{\lambda-\tilde b-ic/2}\right)^{N}\left(\frac{\lambda+\tilde b+ic/2}{\lambda+\tilde b-ic/2}\right)^{N}\left(\frac{\lambda-\tilde d+ic/2}{\lambda-\tilde d-ic/2}\right)\left(\frac{\lambda+\tilde d+ic/2}{\lambda+\tilde d-ic/2}\right)=\Pi_{j\neq k}^{M}\left(\frac{\lambda_j-\lambda_k+ic}{\lambda_j-\lambda_k-ic}\right)\left(\frac{\lambda_j+\lambda_k+ic}{\lambda_j+\lambda_k-ic}\right)\end{equation}

\vspace{2mm}

\begin{equation}\label{Ebae}
e^{-ikL} = \Pi_{j=1}^M\left(\frac{\tilde b+\lambda_j+ic/2}{\tilde b+\lambda_j-ic/2}\right)\left(\frac{\tilde b-\lambda_j+ic/2}{\tilde b-\lambda_j-ic/2}\right)\end{equation}
 These are the equations presented in the main text. Here we have renamed the parameters $\tilde b= bc$, $\tilde d= dc$.

\subsection{Reduced Bethe Equations}

 In order for us to understand the excitation spectrum which consists of holes and complex Bethe roots, we need to reduce the above Bethe equations into a set of constraint equations for the positions of holes and complex Bethe roots \cite{DestriLowenstein}.

\vspace{2mm}

The solutions to the Bethe equations can be categorized into 3 types:

1) Real solutions $\lambda_k$, $k=1,2...M_r$ 

2) Closed pairs $\eta^c_l\pm i\zeta^c_l$, $l=1,2...M_{cp}, \;\beta_l <  c$

3) Wide pairs $\eta^w_r\pm i\zeta^w_r$, $r=1,2...M_{wp}, \; \delta_r>c$

\vspace{2mm}

The Bethe equations are now expressed in terms of the above three different types of solutions and then the density distribution of the real $\lambda$ solutions is found in the presence of the holes (which correspond to the omitted $\lambda_j$ in the ground state sequence $I^0$) and the above complex Bethe roots. The obtained density distribution is then used to integrate out the real $\lambda$ solutions from the Bethe equations  which is equivalent to integrating out the Fermi sea. This transforms the Bethe equations into the Reduced Bethe equations.

The Bethe equations for the real solutions are:

$$\left(\frac{\lambda-\tilde b+ic/2}{\lambda-\tilde b-ic/2}\right)^{N}\left(\frac{\lambda+\tilde b+ic/2}{\lambda+\tilde b-ic/2}\right)^{N}\left(\frac{\lambda-\tilde d+ic/2}{\lambda-\tilde d-ic/2}\right)\left(\frac{\lambda+\tilde d+ic/2}{\lambda+\tilde d-ic/2}\right)=\Pi_{j\neq k}^{M}\left(\frac{\lambda_j-\lambda_k+ic}{\lambda_j-\lambda_k-ic}\right)\left(\frac{\lambda_j+\lambda_k+ic}{\lambda_j+\lambda_k-ic}\right)$$
$$\times \Pi_{l}^{M_{cp}}\left(\frac{\lambda_j-\eta^c_l+i(c-\zeta^c_l)}{\lambda_j-\eta^c_l-i(c-\zeta^c_l)}\right)\left(\frac{\lambda_j-\eta^c_l+i(c+\zeta^c_l)}{\lambda_j-\eta^c_l-i(c+\zeta^c_l)}\right)\left(\frac{\lambda_j+\eta^c_l+i(c-\zeta^c_l)}{\lambda_j+\eta^c_l-i(c-\zeta^c_l)}\right)\left(\frac{\lambda_j+\eta^c_l+i(c+\zeta^c_l)}{\lambda_j+\eta^c_l-i(c+\zeta^c_l)}\right)$$
\begin{equation}\label{Rbaer}
\times \Pi_{r}^{M_{wp}}\left(\frac{\lambda_j-\eta^w_r-i(\zeta^w_r-c)}{\lambda_j-\eta^w_r+i(\zeta^w_r-c)}\right)\left(\frac{\lambda_j-\eta^w_r+i(\zeta^w_r+c)}{\lambda_j-\eta^w_r-i(\zeta^w_r+c)}\right)\left(\frac{\lambda_j+\eta^w_r-i(\zeta^w_r-c)}{\lambda_j+\eta^w_r+i(\zeta^w_r-c)}\right)\left(\frac{\lambda_j+\eta^w_r+i(\zeta^w_r+c)}{\lambda_j+\eta^w_r-i(\zeta^w_r+c)}\right).\end{equation}

\vspace{3mm}

Bethe equations for the closed pairs are:
$$\left(\frac{\eta^c+\tilde b+i(\zeta^c+c/2)}{\eta^c+\tilde b+i(\zeta^c-c/2)}\right)^{N}\left(\frac{\eta^c-\tilde b+i(\zeta^c+c/2)}{\eta^c-\tilde b+i(\zeta^c-c/2)}\right)^{N}\left(\frac{\eta^c+\tilde d+i(\zeta^c+c/2)}{\eta^c+\tilde d+i(\zeta^c-c/2)}\right)\left(\frac{\eta^c-\tilde d+i(\zeta^c+c/2)}{\eta^c-\tilde d+i(\zeta^c-c/2)}\right)=$$$$\Pi_k^{M_r}\left(\frac{\eta^c-\lambda_k+i(c+\zeta^c)}{\eta^c-\lambda_k-i(c-\zeta^c)}\right)\left(\frac{\eta^c+\lambda_k+i(c+\zeta^c)}{\eta^c+\lambda_k-i(c-\zeta^c)}\right)$$
$$\times\Pi_l^{M_{cp}}\left(\frac{\eta^c-\eta^c_l+i(c+\zeta^c-\zeta^c_l)}{\eta^c-\eta^c_l-i(c-\zeta^c+\zeta^c_l)}\right)\left(\frac{\eta^c-\eta^c_l+i(c+\zeta^c+\zeta^c_l)}{\eta^c-\eta^c_l-i(c-\zeta^c-\zeta^c_l)}\right)\left(\frac{\eta^c+\eta^c_l+i(c+\zeta^c-\zeta^c_l)}{\eta^c+\eta^c_l-i(c-\zeta^c+\zeta^c_l)}\right)\left(\frac{\eta^c+\eta^c_l+i(c+\zeta^c+\zeta_l)}{\eta^c+\eta^c_l-i(c-\zeta^c-\zeta^c_l)}\right)$$
\begin{equation}\label{Rbaec}
\times\Pi_r^{M_{wp}}\left(\frac{\eta^c-\eta^w_r+i(\zeta^c-\zeta^w_r+c)}{\eta^c-\eta^w_r+i(\zeta^c-\zeta^w_r-c)}\right)\left(\frac{\eta^c-\eta^w_r+i(\zeta^c+\zeta^w_r+c)}{\eta^c-\eta^w_r+i(\zeta^c+\zeta^w_r-c)}\right)\left(\frac{\eta^c+\eta^w_r+i(\zeta^c-\zeta^w_r+c)}{\eta^c+\eta^w_r+i(\zeta^c-\zeta^w_r-c)}\right)\left(\frac{\eta^c+\eta^w_r+i(\zeta^c+\zeta^w_r+c)}{\eta^c+\eta^w_r+i(\zeta^c+\zeta^w_r-c)}\right).\end{equation}

\vspace{3mm}

Bethe equations for the wide pairs are:
$$\left(\frac{\eta^w+\tilde b+i(\zeta^w+c/2)}{\eta^w+\tilde b+i(\zeta^w-c/2)}\right)^{N}\left(\frac{\eta^w-\tilde b+i(\zeta^w+c/2)}{\eta^w-\tilde b+i(\zeta^w-c/2)}\right)^{N}\left(\frac{\eta^w+\tilde d+i(\zeta^w+c/2)}{\eta^w+\tilde d+i(\zeta^w-c/2)}\right)\left(\frac{\eta^w-\tilde d+i(\zeta^w+c/2)}{\eta^w-\tilde d+i(\zeta^w-c/2)}\right)=$$$$\Pi_k^{M_r}\left(\frac{\eta^w-\lambda_k+i(c+\zeta^w)}{\eta^w-\lambda_k-i(c-\zeta^w)}\right)\left(\frac{\eta^c+\lambda_k+i(c+\zeta^w)}{\eta^c+\lambda_k-i(c-\zeta^w)}\right)$$$$\times\Pi_l^{M_{cp}}\left(\frac{\eta^w-\eta^c_l+i(\zeta^w-\zeta^c_l+c)}{\eta^w-\eta^c_l+i(\zeta^w-\zeta^c_l-c)}\right)\left(\frac{\eta^w-\eta^c_l+i(\zeta^w+\zeta^c_l+c)}{\eta^w-\eta^c_l+i(\zeta^w+\zeta^c_l-c)}\right)\left(\frac{\eta^w+\eta^c_l+i(\zeta^w-\zeta^c_l+c)}{\eta^w+\eta^c_l+i(\zeta^w-\zeta^c_l-c)}\right)\left(\frac{\eta^w+\eta^c_l+i(\zeta^w+\zeta_l+c)}{\eta^w+\eta^c_l+i(\zeta^w+\zeta^c_l-c)}\right)$$
\begin{equation}\label{Rbaew}
\times\Pi_r^{M_{wp}}\left(\frac{\eta^w-\eta^w_r+i(\zeta^w-\zeta^w_r+c)}{\eta^w-\eta^w_r+i(\zeta^w-\zeta^w_r-c)}\right)\left(\frac{\eta^w-\eta^w_r+i(\zeta^w+\zeta^w_r+c)}{\eta^w-\eta^w_r+i(\zeta^w+\zeta^w_r-c)}\right)\left(\frac{\eta^w+\eta^w_r+i(\zeta^w-\zeta^w_r+c)}{\eta^w+\eta^w_r+i(\zeta^w-\zeta^w_r-c)}\right)\left(\frac{\eta^w+\eta^w_r+i(\zeta^w+\zeta^w_r+c)}{\eta^w+\eta^w_r+i(\zeta^w+\zeta^w_r-c)}\right).\end{equation}

Applying logarithm to \eqref{Rbaer} we obtain

$$\sum_{\sigma=\pm}N \Theta(\lambda_j+\sigma \tilde b,c/2)+\Theta(\lambda_j+\sigma \tilde d,c/2)+\Theta\left(2\lambda_j,c\right)=\pi I_j + \sum_{\sigma=\pm}\sum_l^{M} \Theta\left(\lambda_j+\sigma\lambda_l,c\right)$$\begin{equation}+\sum_l^{M_{cp}} \left[ \Theta\left(  \lambda_j+\sigma\eta^c_l,c-\zeta^c_l\right)+ \Theta\left( \lambda_j+\sigma\eta^c_l,c+\zeta^c_l\right)\right] +\sum_r^{M_{wp}} \left[ \Theta\left(\lambda_j+\sigma\eta^w_r,\zeta^w_r+c\right)-\Theta\left(\lambda_j+\sigma\eta^w_r,\zeta^w_r-c\right)\right]. \end{equation}
 Differentiating with respect to $\lambda_j$ we obtain,
\begin{eqnarray}\label{LogRbae}\rho(\lambda)+\frac{1}{2}\delta(\lambda)+\sum_j^n\frac{1}{2}\delta(\lambda-\theta_j)+\frac{1}{2}\delta(\lambda+\theta_j)+ \int \rho(\mu) \varphi(\lambda-\mu,c)= \varphi(2\lambda,c)+ \sum_{\sigma=\pm}\{N\varphi(2\lambda+2\sigma\tilde b,c)+\varphi(2\lambda+\sigma\tilde d,c)\\
-\sum_{l=1}^{M_{cp}}\frac{1}{2}\left[\varphi(\lambda+\sigma \eta_l^c,c-\zeta^c_l)+\varphi(\lambda+\sigma \eta_l^c,c+\zeta^c_l)\right]-\sum_{r=1}^{M_{wp}}\frac{1}{2}\left[\varphi(\lambda+\sigma\eta^w_r,\zeta^w_r+c)-\varphi(\lambda+\sigma\eta^w_r,\zeta^w_r-c)\right]\}\end{eqnarray}

Where we have added $n$ holes at positions $\theta_j$, $j=1...n$. Taking the Fourier transformation we obtain,
\begin{eqnarray}\label{Drbae}\tilde\rho(\omega)=\frac{1}{4\sqrt{2\pi}}\frac{1}{\cosh[\frac{c\omega}{2}]}\times(2N\cos[\tilde b \omega]+2\cos[\tilde d \omega]+1-e^{\frac{c|\omega|}{2}}-\sum_{j=1}^{n}2\cos[\theta_j\omega]e^{c|\omega|/2}\\-4\sum_{l=1}^{M_{cp}}\cos[\eta^c_l\omega]\cosh[\zeta^c_l\omega]e^{-c|\omega |/2}+4\sum_{r=1}^{M_{wp}}\cos[\eta^w_r\omega]\sinh[c|\omega|]e^{-(\zeta^w_r-c/2)|\omega |}).\end{eqnarray}
The value of the Fourier transformed density at the origin gives the following relation,
\begin{equation}\label{Rcount}M=M_r+2M_{cp}+2M_{wp}= \frac{N+1}{2}-\frac{n}{2}+M_{cp}+2M_{wp},\end{equation}
from which we get \begin{equation}\label{Rspin}s^z=\frac{N+1}{2}-M=\frac{n}{2}-M_{cp}-2M_{wp}.\end{equation}

Consider the following term in the Bethe equations corresponding to wide pairs \eqref{Rbaew},
 \begin{eqnarray}\nonumber\Pi_k^{M_r}\left(\frac{\eta^w-\lambda_k+i(c+\zeta^w)}{\eta^w-\lambda_k-i(c-\zeta^w)}\right)\left(\frac{\eta^c+\lambda_k+i(c+\zeta^w)}{\eta^c+\lambda_k-i(c-\zeta^w)}\right)\\=\exp\left\{\int \mathrm{d}\lambda \;\rho(\lambda) \left[\log\left(\frac{i(\lambda-\eta^w)+(\zeta^w+c)}{i(\lambda-\eta^w)+(\zeta^w-c)}\right)+\log\left(\frac{i(\lambda+\eta^w)-(\zeta^w+c)}{i(\lambda+\eta^w)-(\zeta^w-c)}\right)\right]\right\}.\end{eqnarray}

Consider the integral in the exponential

\begin{equation}\label{Rbaeexp}
\mathcal{J}=\int d\lambda \;\rho(\lambda) \left[\log\left(\frac{i(\lambda-\eta^w)+(\zeta^w+c)}{i(\lambda-\eta^w)+(\zeta^w-c)}\right)+\log\left(\frac{i(\lambda+\eta^w)-(\zeta^w+c)}{i(\lambda+\eta^w)-(\zeta^w-c)}\right)\right].\end{equation}
To perform the integral we use the following formulas
$$\int d\lambda \; f(\lambda) \;\log\left[\frac{i(\lambda-\xi)+a}{i(\lambda-\xi)+b}\right]=\sqrt{2\pi}\int_0^{\infty}\frac{d\omega}{\omega}\tilde f(\omega) e^{i\xi\omega}(e^{-b\omega}-e^{-a\omega}),$$
\begin{equation}\label{For}
\int d\lambda \; f(\lambda) \;\log\left[\frac{i(\lambda-\xi)+a}{i(\lambda-\xi)-b}\right]=\sqrt{2\pi}\int_0^{\infty}\frac{d\omega}{\omega}\; (\tilde f(-\omega)e^{-b\omega-i\xi\omega}-\tilde f(\omega)e^{-a\omega+i\xi\omega}))-i\pi\tilde f(0).\end{equation}
where $f(\lambda)$ is a real valued function and $a,b>0$.
Using these formulas in \eqref{Rbaeexp}, we get
\begin{equation}\mathcal{J}=2\sqrt{2\pi}\int_0^{\infty}\left(\frac{1}{\omega}\right)\;e^{i\eta^w\omega}e^{-\zeta^w\omega}\sinh\left[c\;\omega\right]\;(\tilde\rho(\omega)+\tilde\rho(-\omega)).\end{equation}

Using the form of $\tilde\rho(\omega)$ \eqref{Drbae} and using the above formulas \eqref{For} again with $\tilde f(\omega)=1$, we obtain
\begin{eqnarray}\nonumber\exp\left\{\mathcal{J}\right\}= \left(\frac{\eta^w+\tilde b+i(\zeta^w+c/2)}{\eta^w+\tilde b+i(\zeta^w-c/2)}\right)^{N}\left(\frac{\eta^w-\tilde b+i(\zeta^w+c/2)}{\eta^w-\tilde b+i(\zeta^w-c/2)}\right)^{N}\left(\frac{\eta^w+\tilde d+i(\zeta^w+c/2)}{\eta^w+\tilde d+i(\zeta^w-c/2)}\right)\left(\frac{\eta^w-\tilde d+i(\zeta^w+c/2)}{\eta^w-\tilde d+i(\zeta^w-c/2)}\right) \\\nonumber
\left(\frac{\eta^w+i(\zeta^w+c/2)}{\eta^w+i(\zeta^w-c/2)}\right)
\left(\frac{\eta^w+i(\zeta^w-c)}{\eta^w+i\zeta^w}\right)\Pi_{j=1}^n\left(\frac{\eta^w+\theta_j+i(\zeta^w-c)}{\eta^w+\theta_j+i\zeta^w}\right)\left(\frac{\eta^w-\theta_j+i(\zeta^w-c)}{\eta^w-\theta_j+i\zeta^w}\right)\\\nonumber \Pi_{l=1}^{M_{cp}}\left(\frac{\eta^c_l+\eta^w+i(\zeta^w-\zeta^c_l)}{\eta^c_l+\eta^w+i(\zeta^w-\zeta^c_l+c)}\right)\left(\frac{\eta^c_l-\eta^w-i(\zeta^w-\zeta^c_l)}{\eta^c_l-\eta^w-i(\zeta^w-\zeta^c_l+c)}\right)\left(\frac{\eta^c_l+\eta^w+i(\zeta^w+\zeta^c_l)}{\eta^c_l+\eta^w+i(\zeta^w+\zeta^c_l+c)}\right)\left(\frac{\eta^c_l-\eta^w-i(\zeta^w+\zeta^c_l)}{\eta^c_l-\eta^w-i(\zeta^w+\zeta^c_l+c)}\right)\\\nonumber\Pi_{r=1}^{M_{wp}}\left(\frac{\eta^w+\eta^w_r+i(\zeta^w+\zeta^w_r)}{\eta^w+\eta^w_r+i(\zeta^w+\zeta^w_r+c)}\right)\left(\frac{\eta^w+\eta^w_r+i(\zeta^w+\zeta^w_r-c)}{\eta^w+\eta^w_r+i(\zeta^w+\zeta^w_r-2c)}\right)\left(\frac{\eta^w-\eta^w_r+i(\zeta^w+\zeta^w_r)}{\eta^w-\eta^w_r+i(\zeta^w+\zeta^w_r+c)}\right)\left(\frac{\eta^w-\eta^w_r+i(\zeta^w+\zeta^w_r-c)}{\eta^w-\eta^w_r+i(\zeta^w+\zeta^w_r-2c)}\right).\end{eqnarray}
Using this in the Bethe equations corresponding to the wide pair and using the selection rule $\eta^c_l\neq0, \zeta^c_l\neq0$, we obtain
\begin{eqnarray}\nonumber\Pi_{ l=1, \eta^c_l\neq 0, \zeta^c_l\neq 0 }^{M_{cp}}\left(\frac{\eta+\eta^c_l+i(\zeta-\zeta^c_l)}{\eta+\eta^c_l+i(\zeta-\zeta^c_l-c)}\right)\left(\frac{\eta-\eta^c_l+i(\zeta-\zeta^c_l)}{\eta-\eta^c_l+i(\zeta-\zeta^c_l-c)}\right)\left(\frac{\eta+\eta^c_l+i(\zeta+\zeta^c_l)}{\eta+\eta^c_l+i(\zeta+\zeta^c_l-c)}\right)\left(\frac{\eta-\eta^c_l+i(\zeta+\zeta^c_l)}{\eta-\eta^c_l+i(\zeta+\zeta^c_l-c)}\right)\\\nonumber\times
\Pi_{r=1}^{M_{wp}}\left(\frac{\eta+\eta^w_r+i(\zeta+\zeta^w_r)}{\eta+\eta^w_r+i(\zeta+\zeta^w_r-2c)}\right)
\left(\frac{\eta-\eta^w_r+i(\zeta+\zeta^w_r)}{\eta-\eta^w_r+i(\zeta+\zeta^w_r-2c)}\right)\left(\frac{\eta-\eta^w_r+i(\zeta-\zeta^w_r+c)}{\eta-\eta^w_r+i(\zeta-\zeta^w_r-c)}\right)\left(\frac{\eta+\eta^w_r+i(\zeta-\zeta^w_r+c)}{\eta+\eta^w_r+i(\zeta-\zeta^w_r-c)}\right)\\\label{Frbaew}\times\Pi_{j=1}^{n} \left(\frac{\eta+\theta_j+i(\zeta-c)}{\eta+\theta_j+i\zeta}\right)\left(\frac{\eta-\theta_j+i(\zeta-c)}{\eta-\theta_j+i\zeta}\right)=1.~~~~\end{eqnarray}

These are the Reduced Bethe equations corresponding to wide pairs.

\vspace{3mm}

Let us now examine  the following term in the Bethe equations corresponding to the closed pair \eqref{Rbaec},
\begin{equation}\Pi_k^{M_r}\left(\frac{\eta^c-\lambda_k+i(c+\zeta^c)}{\eta^c-\lambda_k-i(c-\zeta^c)}\right)\left(\frac{\eta^c+\lambda_k+i(c+\zeta^c)}{\eta^c+\lambda_k-i(c-\zeta^c)}\right)=\exp\left\{\int d\lambda \; \rho(\lambda) \; \left[\log\left(\frac{\eta^c-\lambda+i(c+\zeta^c)}{\eta^c-\lambda-i(c-\zeta^c)}\right)+\log\left(\frac{\eta^c+\lambda+i(c+\zeta^c)}{\eta^c+\lambda-i(c-\zeta^c)}\right)\right]\right\}.\end{equation}

Consider the integral inside the exponential,
\begin{equation}I=\int d\lambda \; \rho(\lambda) \; \left[\log\left(\frac{\eta^c-\lambda+i(c+\zeta^c)}{\eta^c-\lambda-i(c-\zeta^c)}\right)+\log\left(\frac{\eta^c+\lambda+i(c+\zeta^c)}{\eta^c+\lambda-i(c-\zeta^c)}\right)\right].\end{equation}

Using the formulas \eqref{For}, we get

\begin{equation}I=2\sqrt{2\pi}\int_0^{\infty} \left(\frac{1}{\omega}\right) e^{-c\omega} \sinh[(\zeta^c-i\eta^c)\omega]\;(\tilde\rho(\omega)+\tilde\rho(-\omega)).\end{equation}

Using the form of $\tilde \rho(\omega)$ \eqref{Drbae} and using the above formulas \eqref{For} again with $\tilde f(\omega)=1$, we obtain

$$N\exp\left\{I\right\}=\left(\frac{\eta^c-\tilde b+i(\zeta^c+c/2)}{\eta^c-\tilde b+i(\zeta^c-c/2)}\right)^{N}\left(\frac{\eta^c+\tilde b+i(\zeta^c+c/2)}{\eta^c+\tilde b+i(\zeta^c-c/2)}\right)^{N}\left(\frac{\eta^c-\tilde d+i(\zeta^c+c/2)}{\eta^c-\tilde d+i(\zeta^c-c/2)}\right)\left(\frac{\eta^c+\tilde d+i(\zeta^c+c/2)}{\eta^c+\tilde d+i(\zeta^c-c/2)}\right)\left(\frac{\eta^c+i(\zeta^c+c/2)}{\eta^c+(\zeta^c-c/2)}\right)$$$$\times\left(\frac{1}{\eta^c+i\zeta^c}\right)\left(\frac{1}{(\eta^c-\theta_j+i\zeta^c)(\eta^c+\theta_j+i\zeta^c)}\right)$$
$$\times\Pi_l^{M_{cp}}\left(\frac{1}{(\eta^c_l-\eta^c-i(\zeta^c-\eta^c_l+c))(\eta^c_l-\eta^c-i(\zeta^c+\eta^c_l+c))(\eta^c_l+\eta^c+i(\zeta^c+\eta^c_l+c))(\eta^c_l+\eta^c+i(\zeta^c-\eta^c_l+c))}\right)$$
\begin{equation}\times \Pi_r^{M_{wp}}\left(\frac{\eta^c+\eta^w_r+i(\zeta^r+\zeta^c)}{\eta^c+\eta^w_r-i(\zeta^r-\zeta^c)}\right)\left(\frac{\eta^c-\eta^w_r+i(\zeta^r+\zeta^c)}{\eta^c-\eta^w_r-i(\zeta^r-\zeta^c)}\right)\left(\frac{\eta^c+\eta^w_r-i(\zeta^r-\zeta^c+c)}{\eta^c+\eta^w_r+i(\zeta^r+\zeta^c+c)}\right)\left(\frac{\eta^c-\eta^w_r-i(\zeta^r-\zeta^c+c)}{\eta^c-\eta^w_r+i(\zeta^r+\zeta^c+c)}\right)\times N\exp\left\{I^\prime(\zeta^c)\right\},\end{equation}
where \begin{equation}I^\prime(\zeta^c) = \begin{cases} I^\prime_1(\zeta^c)+I^\prime_2(\zeta^c)+I^\prime_h(\zeta^c)+I^\prime_{cp}(\zeta^c), \hspace{4mm} \zeta^c<c/2 \\       I^{\prime\prime}_1(\zeta^c)+I^{\prime\prime}_2(\zeta^c)+I^{\prime}_h(\zeta^c)+I^{\prime}_{cp}(\zeta^c), \hspace{4mm} \zeta^c>c/2 \end{cases},\end{equation}

\begin{equation}I^\prime_1(\zeta^c)=-\int_0^{\infty} \frac{\left(N\;\sinh[(\zeta^c-i\eta^c+i\tilde b)\omega]+N\;\sinh[(\zeta^c-i\eta^c-i\tilde b)\omega]+\sinh[(\zeta^c-i\eta^c+i\tilde d)\omega]+\sinh[(\zeta^c-i\eta^c-i\tilde d)\omega]\right)}{\omega \; \cosh[\frac{c\omega}{2}]},\end{equation}

\begin{equation}I^{\prime\prime}_1(\zeta^c)=\int_0^{\infty} \frac{\left(N\;\sinh[(\zeta^c-i\eta^c+i\tilde b-c)\omega]+N\;\sinh[(\zeta^c-i\eta^c-i\tilde b-c)\omega]+\sinh[(\zeta^c-i\eta^c+i\tilde d-c)\omega]+\sinh[(\zeta^c-i\eta^c-i\tilde d-c)\omega]\right)}{\omega \; \cosh[\frac{c\omega}{2}]},\end{equation}

\begin{equation}I^{\prime}_2(\zeta^c)=-\int_0^{\infty}\frac{\sinh[(\zeta^c-i\eta^c)\omega]}{\omega\;\cosh[\frac{c\omega}{2}]}-\int_0^{\infty}\frac{\cosh[(\zeta^c-i\eta^c-c/2)\omega]}{\omega\;\cosh[\frac{c\omega}{2}]},\end{equation}

\begin{equation} I^{\prime\prime}_2(\zeta^c)=\int_0^{\infty}\frac{\sinh[(\zeta^c-i\eta^c-c)\omega]}{\omega\;\cosh[\frac{c\omega}{2}]}-\int_0^{\infty}\frac{\cosh[(\zeta^c-i\eta^c-c/2)\omega]}{\omega\;\cosh[\frac{c\omega}{2}]},\end{equation}

\begin{equation}I^{\prime}_h(\zeta^c)=\sum_{j=1}^{n}\int_0^{\infty}\frac{\cosh[(\zeta^c-c/2-i(\eta^c-\theta_j))\omega]+\cosh[(\zeta^c-c/2-i(\eta^c+\theta_j))\omega]}{\omega\;\cosh[\frac{c\omega}{2}]},\end{equation}

\begin{eqnarray}I^{\prime}_{cp}&=&-\sum_{l=1}^{M_{cp}}\int_0^{\infty} \frac{e^{-c\omega}}{\omega\;\cosh[\frac{c\omega}{2}]}\left(\cosh[(\zeta^c-\zeta^c_l-c/2+i(\eta^c_l-\eta^c))\omega]+\cosh[(\zeta^c+\zeta^c_l-c/2+i(\eta^c_l-\eta^c))\omega]\right.\\&&\left.+\cosh[(\zeta^c-\zeta^c_l-c/2-i(\eta^c_l+\eta^c))\omega]+\cosh[(\zeta^c+\zeta^c_l-c/2-i(\eta^c_l+\eta^c))\omega]\right). \end{eqnarray}
Using this in the Bethe equations corresponding to the closed pair \eqref{Rbaec}, we get
$$F(\zeta^c)\equiv \exp\left\{I^\prime(\zeta^c)\right\}\left(\frac{1}{\eta^c+i\zeta^c}\right)\left(\frac{1}{(\eta^c-\theta_j+i\zeta^c)(\eta^c+\theta_j+i\zeta^c)}\right) $$$$\times \Pi_{l=1}^{M_{cp}}\left(\frac{1}{(\eta^c-\eta^c_l+i(\zeta^c-\zeta^c_l-c))(\eta^c-\eta^c_l+i(\zeta^c+\zeta^c_l-c))(\eta^c+\eta^c_l+i(\zeta^c+\zeta^c_l-c))(\eta^c+\eta^c_l+i(\zeta^c-\zeta^c_l-c))}\right)$$
\begin{equation}\Pi_{r=1}^{M_{wp}} \left(\frac{\eta^c+\eta^w_r+i(\zeta^w_r+\zeta^c)}{\eta^c+\eta^w_r-i(\zeta^w_r-\zeta^c)}\right)\left(\frac{\eta^c-\eta^w_r+i(\zeta^w_r+\zeta^c)}{\eta^c-\eta^w_r-i(\zeta^w_r-\zeta^c)}\right)\left(\frac{\eta^c-\eta^w_r+i(\zeta^c-\zeta^w_r+c)}{\eta^c-\eta^w_r+i(\zeta^c+\zeta^w_r-c)}\right)\left(\frac{\eta^c+\eta^w_r+i(\zeta^c-\zeta^w_r+c)}{\eta^c+\eta^w_r+i(\zeta^c+\zeta^w_r-c)}\right)=1.\end{equation}

\vspace{2mm}

To obtain the final form of the reduced Bethe equations corresponding to closed pairs we need to eliminate the exponential factor $\exp\left\{I^\prime(\zeta^c)\right\}$ in the above equation. This can be achieved by dividing $F(\zeta^c)$ by $F^*(c-\zeta^c)$ and noting that $I^*(c-\zeta^c)=I(\zeta^c)$. Where $^*$ denotes complex conjugation. We have,

$$\frac{F(\zeta^c)}{F^*(c-\zeta^c)}\equiv\Pi_{j=1}^{n} \left(\frac{\eta+\theta_j+i(\zeta-c)}{\eta+\theta_j+i\zeta}\right)\left(\frac{\eta-\theta_j+i(\zeta-c)}{\eta-\theta_j+i\zeta}\right)\times$$$$\Pi_{ l=1, \eta^c_l\neq 0, \zeta^c_l\neq 0}^{M_{cp}}\left(\frac{\eta+\eta^c_l+i(\zeta-\zeta^c_l)}{\eta+\eta^c_l+i(\zeta-\zeta^c_l-c)}\right)\left(\frac{\eta-\eta^c_l+i(\zeta-\zeta^c_l)}{\eta-\eta^c_l+i(\zeta-\zeta^c_l-c)}\right)\left(\frac{\eta+\eta^c_l+i(\zeta+\zeta^c_l)}{\eta+\eta^c_l+i(\zeta+\zeta^c_l-c)}\right)\left(\frac{\eta-\eta^c_l+i(\zeta+\zeta^c_l)}{\eta-\eta^c_l+i(\zeta+\zeta^c_l-c)}\right)\times$$\begin{equation}\label{Frbaec}\Pi_{r=1}^{M_{wp}}\left(\frac{\eta+\eta^w_r+i(\zeta+\zeta^w_r)}{\eta+\eta^w_r+i(\zeta+\zeta^w_r-2c)}\right)
\left(\frac{\eta-\eta^w_r+i(\zeta+\zeta^w_r)}{\eta-\eta^w_r+i(\zeta+\zeta^w_r-2c)}\right)\left(\frac{\eta-\eta^w_r+i(\zeta-\zeta^w_r+c)}{\eta-\eta^w_r+i(\zeta-\zeta^w_r-c)}\right)\left(\frac{\eta+\eta^w_r+i(\zeta-\zeta^w_r+c)}{\eta+\eta^w_r+i(\zeta-\zeta^w_r-c)}\right)=1.\end{equation}

\vspace{2mm}

Here, the selection rule $\eta^c_l\neq0, \zeta^c_l\neq0$ was applied. The reduced Bethe equations corresponding to wide pairs \eqref{Frbaew} are exactly same as those corresponding to closed pairs \eqref{Frbaec}.

For the unscreened phase, the calculation can be done by the same method described above. After applying the appropriate selection rules, we obtain the same exact reduced Bethe equations as that of the screened phase \eqref{Frbaew}, \eqref{Frbaec}.
In this case, an additional selection rule which disallows the solution $\lambda=\pm i(\tilde a-c/2)$ as a closed pair is applied. This selection rule is self imposed by the resulting reduced Bethe equations in the unscreened phase. This solution might although exist as a wide pair for $\tilde a>3c/2$. From \eqref{Rcount} we see that four holes are needed for the existence of this boundary string and the resulting excitation would be a singlet \eqref{Rspin}. Here we solve for the possible positions of the four holes and find that no solutions exist, which means that this boundary string cannot be present in the system as a singlet excitation. The reduced Bethe equations in the presence of this boundary string and four holes at positions $\theta_j, j=1..4$ are

\begin{equation}\Pi_{j=1}^{4} \frac{\theta_j^2+(\tilde a-3c/2)^2}{\theta_j^2+(\tilde a-c/2)^2}=1\end{equation}

Let us fix the positions of 3 holes $\theta_2, \theta_3, \theta_4$ and solve for $\theta_1$. We have

\begin{eqnarray}\left(1-\Pi_{j=2}^{3} \frac{\theta_j^2+(\tilde a-c/2)^2}{\theta_j^2+(\tilde a-3c/2)^2}\right)\theta_1^2=(\tilde a-c/2)^2\left(\Pi_{j=2}^{3} \frac{\theta_j^2+(\tilde a-c/2)^2}{\theta_j^2+(\tilde a-3c/2)^2}\right)-(\tilde a-3c/2)^2\end{eqnarray}

The term inside the brackets on the left hand side is always negative. The right hand side is always positive. This means $\theta_1^2$ is negative hence we cannot have any real solutions for $\theta_1$. This implies we cannot have the boundary string as there are no real solutions for the holes.

\subsection{Elementary Excitations}
In order to study the effect of a magnetic field on the system and thereby determine the existence of a Kondo effect as well as understand the transition from the screened to the unscreened phases we need to study the low energy excitations of the system. In this section we constuct the triplet and singlet spinon excitations.

\textit{Triplet Excitation:} The triplet excitation is formed by adding two holes to the ground state of the Bethe roots.  Adding two holes at $\theta_1$ and $\theta_2$, we obtain from \eqref{Drbae}
\begin{eqnarray}
\tilde\rho(\omega)=\tilde\rho_{\text{gs}}(\omega)+\delta\tilde\rho^h(\omega),
\end{eqnarray}
where $\tilde{\rho}_\text{gs}$ refers to the ground state distribution either in the screened \eqref{scr} or unscreened \eqref{unscr} phase and 
\begin{equation}\delta\tilde\rho^h(\omega)=-\frac{1}{2\sqrt{2\pi}}\frac{\cos[\theta_1 \omega]+\cos[\theta_2 \omega]}{\cosh[\frac{c\omega}{2}]}e^{\frac{c|\omega|}{2}}\end{equation}
is the shift in the distribution caused by the holes. Using the same method as in the ground state we can calculate that the spin of this excitation is $S_z = 1$. Justifying the moniker of triplet excitation. 
It is also worthwhile to calculate the energy. Using $E=\sum_{j=1}^Nk_j$ along with \eqref{logEnergy} and replacing the sum over Bethe roots by an integral over $\rho(\lambda)$ we find that  energy of the the triplet excitation is 
\begin{eqnarray}
\delta E= D \int \delta \rho^h(\lambda)\sum_{\sigma=\pm} \Theta(\tilde b+\sigma\lambda,c/2)
\end{eqnarray}
which yields, 
\begin{eqnarray}\label{tripletenergy}\delta E=  \sum_{l=1}^2D\arctan\left[\frac{\cosh\left(\frac{\theta_l\pi}{c}\right)}{\sinh\left(b\pi\right)}\right].\end{eqnarray}
Therefore the total excitation energy is a sum of two terms one from each hole. We interpret this as two spinons, each carrying spin $1/2$ which are symmetrically coupled in the triplet state. 
In addition to this we see that in contrast to the gapless, linear dispersion of the bare left and right movers of the system, the spinons have acquired a gap. The minimum energy of a single spinon occurs at $\theta_l=0$ and it is given by
\begin{equation}\Delta= D\arctan\left[\frac{1}{\sinh\left(b\pi\right)}\right]\end{equation}
This is the same dynamical mass generation present in the chiral Gross-Neveu model. Note that the mass vanishes when $g_1\to 0$.

\textit{Singlet Excitation}:  In addition to the triplet excitation there must also be a singlet. This is created by adding two holes as before but also introducing a complex conjugate pair of Bethe roots called a 2-string, $\lambda_s \pm ic/2$, which belongs to the closed pair category in \eqref{Frbaec}. We obtain from \eqref{Drbae},

\begin{eqnarray}
\tilde\rho(\omega)=\tilde\rho_\text{gs}(\omega)+\delta\tilde\rho^h(\omega)+\delta\tilde\rho^s(\omega)
\end{eqnarray}
where $\Delta\tilde\rho^h$ is the same as in before and the contribution from the string is 
\begin{equation}\delta\tilde\rho^s(\omega)=-\frac{1}{\sqrt{2\pi}}e^{-\frac{c|\omega|}{2}}\cos[\lambda_s\omega]\end{equation}
 The position of the 2-string $\lambda_s$ depends on the positions of the holes, this can be calculated by using the reduced Bethe equations. In the presence of two holes at $\theta_1$ and $\theta_2$ and one closed pair with imaginary part $c/2$, the reduced Bethe equations \eqref{Frbaec} take the form

$$\left(\frac{i+\frac{2(\lambda_s+\theta_1)}{c}}{i-\frac{2(\lambda_s+\theta_1)}{c}}\frac{i+\frac{2(\lambda_s-\theta_1)}{c}}{i-\frac{2(\lambda_s-\theta_1)}{c}}\frac{i+\frac{2(\lambda_s+\theta_2)}{c}}{i-\frac{2(\lambda_s+\theta_2)}{c}}\frac{i+\frac{2(\lambda_s-\theta_2)}{c}}{i-\frac{2(\lambda_s-\theta_2)}{c}}\right)=1,$$

which gives $\lambda_s= \pm\sqrt{\frac{\theta_1^2}{2}+\frac{\theta_2^2}{2}+\frac{c^2}{4}}$. 

\vspace{2mm}

 This result which is in agreement with the result obtained in \cite{GrisaruMezincescNepomechie}, is different to the periodic case where the position of the string is at the midpoint of the two holes. $\lambda_s=0$ is also a solution to the above equation and it is a trivial solution which occurs for any number of holes. This is not a valid solution since for more than two holes, the reduced Bethe equations yield irregular number of other complex solutions when solved in the presence of this trivial solution, hence it should be discarded. 

The spin of this excitation is calculated along similar lines to before. Taking into account the extra 2-string we have that  $S_z=(N+1)/2-2-\sqrt{2\pi} \tilde\rho(0)=0$. Thus the the excitation is a singlet and moreover performing the same calculation of the energy we find that $\delta\rho^s$ is enough to completely cancel the bare energy of adding the 2-string. Therefore the energy is also given by \eqref{tripletenergy} making it degenerate with the triplet excitation. In the scaling limit the energy becomes
\begin{eqnarray}
\delta E=\Delta\cosh{\left(\frac{\pi\theta_1}{c}\right)}+\Delta\cosh{\left(\frac{\pi\theta_2}{c}\right)}.
\end{eqnarray}
which shows that  the Lorentz invariance of the model is restored in the scaling limit. 
In the next section we shall see that a similar scaling limit of the impurity parameter will also be necessary. 


\subsection{Magnetization}
We now turn to the calculation of the magnetization in the presence of a small magnetic field, $h$. To do this we add a term $-\mu h S_z$ to the Hamiltonian, where $S_z=\frac{1}{2} \sum_{i=0}^{N} \sigma_i^z$ and so we are looking for the ground state of the following Hamiltonian
\begin{equation}\label{Hprime}H'=H -\mu h S_z\end{equation}
which shall provide us with the magnetization of the system. 
Since $H$ commutes with $S^z$, the eigenstates of $H'$ are same as that of $H$ however the energy levels are shifted. Consequently the ground state will be different and will require a balancing of the energy contribution of both terms in $H'$. We already know that the for $h=0$ involves all the electrons forming a singlet and the impurity being either screened or unscreened, in contrast the magnetic field term is minimized by maximizing the total number of up spins, i.e. $S_z=N/2$. Thus in the presence of a magnetic field minimizing the energy requires that a number of holes be included in the ground state.

Since there is an energy gap in the system, we will find that there exists a critical field up until which the bulk will not magnetize. This critical field naturally depends on the energy gap and can be expected to be $h_c=2\Delta/\mu$ corresponding to the minimum energy required to create a triplet excitation. When the applied magnetic field is greater than the critical value, hole excitations occur symmetrically about $\theta=0$ up until $\theta=\pm B(h)$ owing to the spectrum being symmetric in $\lambda$ , where $B(h)$ depends on the applied magnetic field \cite{BahderRezayiSak}. 

Let the density of the holes be denoted by $\rho^h(\lambda), \hspace{2mm} |\lambda| < B(h) $ and $\rho(\lambda) =0 , \hspace{2mm} |\lambda| < B(h)$ We combine these and define a new function $\rho_B(\lambda)= \rho^h(\lambda) + \rho(\lambda)$ which is the distribution of particles and holes in the ground state. Starting from the Bethe equations we find that it satisfies the following integral equation
\begin{equation}\label{rhoB} h(\lambda)=\rho_{B}(\lambda) + \int_{|\mu|>B} f(\lambda-\mu) \rho_B(\mu),\end{equation}
where $h(\lambda)=g(\lambda)-\frac{1}{2}\delta(\lambda)$. Here $g(\lambda)$ refers to ether $g_s(\lambda)$ or $g_u(\lambda)$.

This equation reduces to the previous equation \eqref{gsdensity} when $B(h)=0$ i.e. for $h=0$. In terms of this the total energy of the system as a function of the magnetization is 
\begin{eqnarray}\label{EnergyB}
E= \frac{D}{L} \int_{|\lambda|>B}\rho_B(\lambda)\sum_{\sigma=\pm} \Theta(\tilde b+\sigma \lambda,c/2)+ \left(\pi D-\mu h\right)S_z.~
\end{eqnarray}
Furthermore, we can write the total spin in terms of $\rho_B(\lambda)$,
\begin{equation}\label{SzB} S_z^\text{s}=\frac{1}{2} \int_{-B}^{B} \rho_B(\lambda) d\lambda, \hspace{4mm} 
S_{z}^\text{us}= \frac{1}{2}+\frac{1}{2} \int_{-B}^{B} \rho_B(\lambda) d\lambda. \end{equation}
where $S_z^\text{u}$ and $S_z^\text{us}$ are the spin in the screened and  unscreened phases. 

We need to  solve \eqref{rhoB} subject to the constraint that \eqref{EnergyB} is minimized after which the magnetization can be determined. Since  
the minimum value of the spin $S_z$ occurs for $B=0$ and the maximum value occurs for $B=\infty$, we can solve \eqref{rhoB} perturbatively in the small parameter $B$ which corresponds to small magnetic field, $h$.  This  is used in \eqref{SzB} to obtain a relation between the spin and the parameter $B$, $S_z=S_z(B)$ and subsequently inserted into \eqref{EnergyB} to give $E(B)$. The ground state occurs for a minimum value of $\epsilon(S)= E(S_z)-E(0)$ which gives the us $B(h)$ and therefore $S_z(h)$, the magnetization. We carry this out separately for both the screened and unscreened cases below. In both cases we find that $2\pi B(h)= c \sqrt{(\mu h-2\Delta)/2\Delta}$ from which we find that the critical  field is $h_c=2\Delta/\mu$ in agreement with our previous expectation. 

\vspace{2mm}

Here we derive the magnetization in the screened phase. By adding and subtracting the integral \begin{equation}\int_{-B}^{B} \frac{2c}{\pi} \frac{\rho_B(\mu)}{c^2+(\lambda-\mu)^2}\end{equation} to the equation \eqref{rhoB} with $g(\lambda)=g_s(\lambda)$ and applying the Fourier transformation we get 

\begin{equation}\label{sm1}\tilde \rho_B(\omega)=\tilde\rho^\prime_0(\omega)+\int_{-B}^{B} \frac{1}{\sqrt{2\pi}}\frac{\rho_B(\mu)e^{i\omega\mu}}{1+e^{c|\omega|}},\end{equation}

where \begin{equation}\tilde\rho^\prime_0(\omega)=\frac{1}{4\sqrt{2\pi}} \frac{\left(2N \cos[\tilde b \omega] + 2\cos[\tilde d \omega]+1\right)}{\cosh[\frac{c \omega}{2}]}.\end{equation}

By applying Fourier transformation to \eqref{sm1} we obtain 

\begin{equation}\label{sm2}\rho_B(\lambda)=\rho^\prime_0(\lambda)-\int_{-B}^{B} \rho_B(\mu) R(\lambda-\mu)  \hspace{3mm} , R(x)=-\frac{1}{2\pi}\int_{-\infty}^{\infty}\frac{e^{-i\omega x}}{1+e^{c|\omega |}}.\end{equation}

\vspace{2mm}

For small magnetic field, by expanding $\rho_B(\lambda)$ around zero and performing the integral in equation \eqref{sm2} we obtain

\begin{equation}\rho_B(0)=\rho^\prime_0(0)-2B \rho_B(0)R(0)-\frac{B^3}{3}\left(\rho_B(0)R(0)\right)^{\prime\prime}+... ,\end{equation}

 Which yields
 
 \begin{equation} \label{Exrhob}\rho_B(0)=\rho^\prime_0(0)-2B\rho^\prime_0(0)R(0)+4B^2R(0)^2\rho^\prime_0(0)+O(B^3).\end{equation}

\vspace{2mm}

Consider the first relation in \eqref{SzB}. Expanding $\rho_B(\lambda)$ around zero and then performing the integral we get

\begin{equation}\label{Exsb} S_z^s=\rho_B(0) B+ \rho^{\prime\prime}_B(0) \frac{B^3}{6}+...\end{equation}

Using \eqref{Exrhob} in \eqref{Exsb}, we obtain

\begin{equation}\label{relsb} S_z^s=\rho^\prime_0(0) B-2B^2\rho^\prime_0(0) R(0)+ B^3\left(4 \rho_0(0)R^2(0)+\frac{\rho^{\prime\prime}(0)}{6}\right)+ O(B^4).\end{equation}

Consider, 

\begin{equation}\epsilon(S_z^s)=E(S_z^s)-E(0)=D\sum_{\sigma=\pm}\int_{|\lambda|>B}(\rho_B(\lambda)-\rho^\prime_0(\lambda))\Theta(\tilde b+\sigma\lambda,c/2)+(\pi D-\mu H)S_z^s,\end{equation}

By using \eqref{sm2} and integrating over $\lambda$ in the above equation, we get

\begin{equation}\epsilon(S_z^s)=-D\int_{-B}^{B} \rho_B(\mu) \arctan\left[\frac{\sinh\left(\frac{\pi\tilde b}{c}\right)}{\cosh\left(\frac{\pi\mu}{c}\right)}\right]+\left(\pi D-\mu H\right)S_z^s.\end{equation}

Expanding the above integrand around zero and performing the integral, and using \eqref{relsb} to eliminate $B$ in favor of $S_z^s$, we obtain 

\begin{equation}\epsilon(S_z^s)=-\mu(h-\frac{2\Delta}{\mu}) S_z^s +\frac{D}{3}\left(\frac{\pi^2}{c^2}\right)\frac{1}{[\rho^\prime_0(0)]^2} \frac{\sinh[\pi b]}{\left(\cosh[\pi b]\right)^2}(S_z^s)^3.\end{equation}
  
For $\mu h<2\Delta$, the coefficient of $S_z^s$ is positive and hence the minimum of $\epsilon(S_z^s)$ occurs at $S_z^s=0$. For $\mu H>2\Delta$, minimizing $\epsilon(S_z^s)$, we obtain

 \begin{equation}S_z^s=\left(\frac{c}{\pi}\right)\frac{\cosh[\pi b]}{\left(\sinh[\pi b]\right)^{1/2}}\left(\frac{1}{D}\right)^{1/2}\left(\mu h-2\Delta\right)^{1/2} \rho^\prime(0).\end{equation}
 
 By using the form of $\rho^\prime(0)$ and trading in the bare parameters for the physical scales $\Delta$ and $T_0$ we arrive at \eqref{magnetscr}. Following the same method one can also derive \eqref{magnetunscr}.
 
 \subsection{Derivation of the RG equations}
Within the Bethe ansatz approach to quantum field theory one can recover the leading order RG equations given. In this section we show how this is done and derive \eqref{RG}.

The gap is given by 
\begin{eqnarray}
\Delta=D\arctan{\left[\frac{1}{\sinh{\pi b}}\right]}
\end{eqnarray}
in order to take the universal scaling limit we take $D\to\infty$ and $b\to 0$ such that $\Delta$ is held fixed. Upon doing this we have 
\begin{eqnarray}
\Delta=2De^{-\pi/2g_1}
\end{eqnarray}
where we have replaced  $b\to1/2g_1$. Inverting this relationship tells us how at weak coupling $g_1$ flows as a function of the energy scale,
\begin{equation}\label{RG1}
    g_1(\Lambda)=\frac{1}{\frac{2}{\pi}\log{\left(\frac{2\Lambda}{\Delta}\right)}}
\end{equation}
where $\Lambda$ is some energy scale and $\Delta$ is held fixed. 
From our expression for $d$ we have
\begin{eqnarray}
c=\frac{2b}{b^2-d^2-1}
\end{eqnarray}
where in the unscreened phase we should take $d^2\to -a^2$. Inserting \eqref{RG1} into the above expression we obtain the flow of $J$ at weak coupling
\begin{eqnarray}\label{RG2}
J(\Lambda)=\frac{\frac{1}{\pi}\log{\left(\frac{2\Lambda}{\Delta}\right)}}{\left[\frac{1}{\pi}\log{\left(\frac{2\Lambda}{\Delta}\right)}\right]^2-1-d^2}
\end{eqnarray}
where we have replaced $c\to 2J$. These equations can be most easily analyzed by turning them into a set of coupled differential equations. Differentiating \eqref{RG1} and \eqref{RG2} with respect to $l=\log{\Lambda}$ we arrive at \eqref{RG}.

\end{document}